# Pressuring the low-temperature orthorhombic phase with a non-trivial topological state of $Ru_2Sn_3$ to room temperature

Shan Zhang[1], Q. D. Gibson[2], Wei Yi[1], Jing Guo[1], Zhe Wang[1], Yazhou Zhou[1], Honghong Wang[1], Shu Cai[1], Ke Yang[3], Aiguo Li[3], Qi Wu[1], Robert J Cava[2], Liling Sun[1,4]† & Zhongxian Zhao[1,4]

[1]*Institute of Physics and University of Chinese Academy of Sciences, Beijing 100190, China*
[2]*Department of Chemistry, Princeton University, Princeton, NJ 08544, USA*
[3]*Shanghai Synchrotron Radiation Facilities, Shanghai Institute of Applied Physics, Chinese Academy of Sciences, Shanghai 201204, China*
[4]*Collaborative Innovation Center of Quantum Matter, Beijing, 100190, China*

We report high pressure studies of the structural stability of $Ru_2Sn_3$, a new type of three dimensional topological insulator (3D-TI) with unique quasi-one dimensional Dirac electron states throughout the surface Brillouin zone of its one-atmosphere low temperature orthorhombic form. Our *in-situ* high-pressure synchrotron x-ray diffraction and electrical resistance measurements reveal that upon increasing pressure the tetragonal to orthorhombic shifts to higher temperature. We find that the stability of the orthorhombic phase that hosts the non-trivial topological ground state can be pushed up to room temperature by an applied pressure of ~ 20 GPa. This is in contrast to the commonly known 3D-TIs whose ground state is usually destroyed under pressure. Our results indicate that pressure provides a possible pathway for realizing a room-temperature topological insulating state in $Ru_2Sn_3$.

Topological insulators host a new state of quantum matter that provides a research platform for fundamental physics and technical applications. Theoretical and experimental investigations have shown that three-dimensional topological insulators (3D-TIs) present a gapless surface state, and that virtually all of the known materials display an isotropic or nearly isotropic surface state Dirac cone with linear energy vs. wavevector dispersion [1-9]. Recently, in contrast, a new type of 3D-TI, $Ru_2Sn_3$, has been found whose surface states are highly anisotropic, displaying an almost flat dispersion along certain high-symmetry directions [10], a type of surface electronic structure that is distinct from those of the standard 3D-TIs. It is generally believed that the non-trivial electron states on the surfaces of TIs are protected by either time reversal symmetry or crystal symmetry [1-5, 11-13]. Once that symmetry is destroyed, however, the topological insulator surface electronic states are expected to transform to a different state. [13-18].

Structurally, $Ru_2Sn_3$ belongs to a family of intermetallic compounds with a defect $TiSi_2$-type structure known as Nowotny chimney-ladder phases. The structures are based on the formation of the Ru-Sn chains [10, 19]. At ambient pressure, $Ru_2Sn_3$ has a non-centrosymmetric tetragonal structure for temperatures between 300 K and 200 K, but transforms to a centrosymmetric orthorhombic structure blow 150 K [20]. The structures are shown in Fig.1. Correspondingly, the temperature dependence of the resistivity of $Ru_2Sn_3$ displays a broad crossover from semiconducting-like behavior to metallic-like behavior in the temperature range of 150~200 K; the breadth of the transition is due to the fact that the transition from the tetragonal (T) to orthorhombic

(O) phase is sluggish. The unique star-shaped topological electronic surface state has been observed in the orthorhombic phase [10, 19-23].

Pressure is an ideal method for tuning a system from one state to another without adding chemical complexity, and therefore the high pressure behavior of this new type of 3D-TI $Ru_2Sn_3$ is of current interest. In this study, we demonstrate that the low-temperature orthorhombic phase that displays the non-trivial topological insulating state of $Ru_2Sn_3$ can be tuned to room temperature through the application of pressure.

The single crystals were synthesized by a solid state reaction method described in Ref.10. Pressure was generated by a diamond anvil cell with two opposing anvils sitting on a Be-Cu supporting plate. Diamond anvils with 300 μm flats were used for this study. A non-magnetic rhenium gasket with 100 μm diameter hole was used for different runs of the high-pressure studies. The four-probe method was applied in the *ab* plane of a single crystal of $Ru_2Sn_3$ for all high pressure resistance measurements. The *in-situ* high pressure x-ray diffraction (XRD) experiments were performed at beamline 15U of the Shanghai Synchrotron Radiation Facilities. Diamonds with low birefringence were selected for the experiments. A monochromatic X-ray beam with a wavelength of 0.6199 Å was adopted for all measurements. To maintain the sample in a quasi-hydrostatic pressure environment, NaCl powder and silicon oil were employed as pressure transmitting media for the high-pressure resistance and XRD measurements, respectively. The pressure was determined by the ruby fluorescence method [24]. The crystallographic cell refinements of the X-ray diffraction data were conducted using the program RIETAN-FP [25].

Figure 2a shows electrical resistance as a function of temperature measured at different pressures. It can be seen that there exists a relatively broad resistance peak with a maximum at about 150 K ($T_{BP}$) when the $Rn_2Sn_3$ sample is subjected to a pressure of about 0.05 GPa. This is in agreement with its ambient pressure behavior [10, 23]. Previous studies have shown that this broad temperature-induced resistance peak is associated with the tetragonal to orthorhombic crystal structure phase transition [10, 20-22]. The two characteristic temperatures, $T_1$ and $T_2$ in Fig. 2a, indicate the boundaries of the mixed phase. Upon increasing pressure, we find that the temperature of the peak, $T_{BP}$, shifts monotonically to higher temperature. At pressures above 17.11 GPa, the resistance peak is no longer visible for temperatures of 280 K and lower, and the resistance-temperature curves show a completely metallic-like behavior.

Based on our resistance data, we summarize the pressure dependence of the resistance maximum, $T_{BP}$, in Fig.2b. It is found that $T_{BP}$ increases with pressure and reaches ~280 K at 17.11 GPa. Extrapolation of $T_{BP}$ to ~300 K gives a critical pressure of ~ 20 GPa. These results reveal that the low-temperature orthorhombic phase with the non-trivial topological insulating state is tuned to room temperature by the application of pressure.

To support the transport results and further clarify the effect of pressure on the crystal structure of $Rn_2Sn_3$, we performed *in-situ* high pressure XRD measurements at room temperature (300 K). Figure 3a shows the XRD patterns obtained at various pressures up to 40.26 GPa. All the observed peaks in the XRD patterns below 1.48 GPa can be indexed well within the tetragonal phase (T phase) structure in space group *P-4c2* (No.116). We find that new diffraction peaks emerge when the pressure is increased to 7.25 GPa; these are indicated by green arrows in the figure. Refinements of the diffraction data for the pressure-induced phase reveal that it is the same as the low-temperature ambient pressure orthorhombic phase (O phase) in space group *Pbcn* (No.60) [20]. At 20.79 GPa, the crystal structure of $Ru_2Sn_3$ becomes fully O phase. Figure 3 (b) shows the pressure dependence of the lattice parameters. Defining the lattice parameters of the T phase as *a, b, c* and the O phase as *a′, b′, c′*, it is found that the lattice parameters of $Ru_2Sn_3$ gradually decrease upon increasing pressure below 7 GPa, but split at pressures above 7 GPa, reflecting that the T-O phase transition begins at this pressure point at room temperature. Our refinement results also indicate that the lattice parameters of these two phases obey the relationship *a′=2b, b′=c, and c′=a*. The pressure dependence of volume is plotted in Fig. 3 (c). Clearly, there are three regions for the pressure-induced structural evolution: the T phase is stable up to 7 GPa (the left region); the T and the O phases coexist for pressures ranging from 7 GPa to 20 GPa (the middle region); and, finally, a single O phase appears at 20 GPa and persists up to 40 GPa, the limit of our measurements (the right region). Figures 3d-3f show the details of the refinement results for the T phase, the T-O mixed phase and the O phase regions. Thus the high pressure XRD results further confirm the existence of a pressure-induced T-O phase transition at room temperature, and support our high pressure resistance measurements.

Finally, we present the pressure-temperature phase diagram for $Ru_2Sn_3$, shown in Fig.4. It is seen that, at ambient pressure, there is a temperature-induced structural transition from the T phase to the O phase in a broad temperature range. Below 50 K, the sample fully enters the O phase that hosts the non-trivial topological insulating ground state (green region). Upon increasing pressure, the T-O phase transition temperature increases, and reaches room temperature at ~ 20 GPa. These results lead us to propose that the unique electronic state of $Ru_2Sn_3$ may be tuned to room temperature by applying pressure, in contrast to the currently known 3D-TIs whose ground state is typically destroyed under pressure [26-28]. Our work indicates that application of pressure may be an avenue for realizing a room temperature topological insulating state in $Ru_2Sn_3$.

**Acknowledgments**

The work in China was supported by the NSF of China (Grants No. 91321207, No. 11427805, No. 11404384, No. U1532267, No. 11604376), the Strategic Priority Research Program (B) of the Chinese Academy of Sciences (Grant No. XDB07020300) and the National Key Research and Development Program of China (Grant No.2016YFA0300300). The work at Princeton was supported by the US NSF MRSEC program, grant DMR-1420541.

†Corresponding authors
llsun@iphy.ac.cn

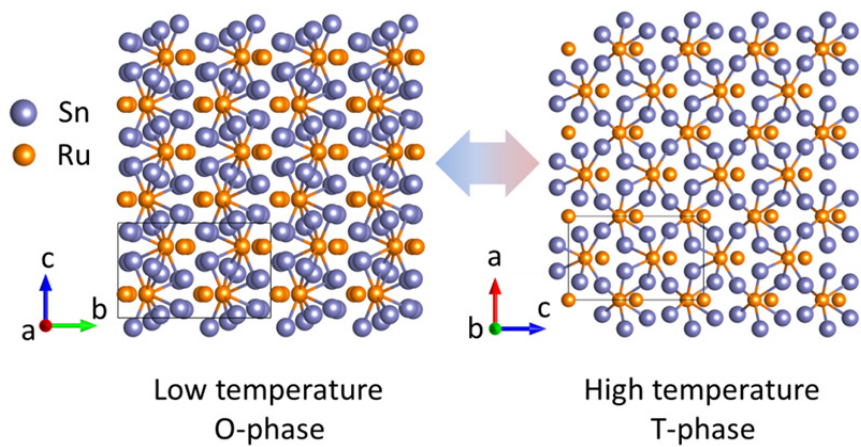

Figure 1 The two crystal structures of $Ru_2Sn_3$. The left panel shows the low-temperature orthorhombic ("O-phase") structure that is the topological insulator, and the right panel shows the high-temperature tetragonal ("T-phase") structure.

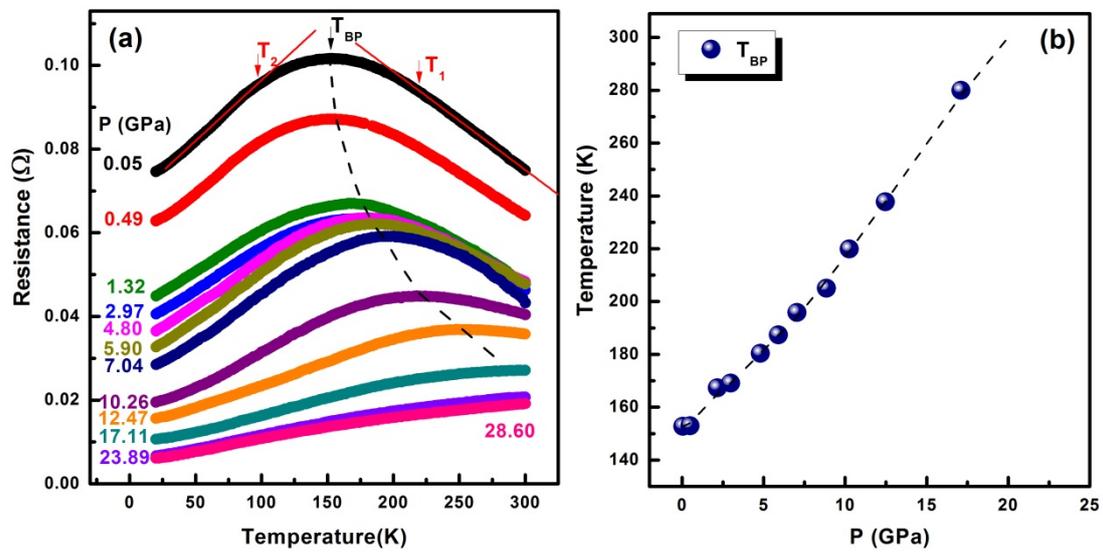

Figure 2 (a) The temperature dependence of the electrical resistance for $Ru_2Sn_3$ measured at different pressures. $T_{BP}$ represents the crossover temperature that signifies the tetragonal-orthorhombic phase transition. (b) Pressure dependence of $T_{BP}$ and its extrapolation to room temperature.

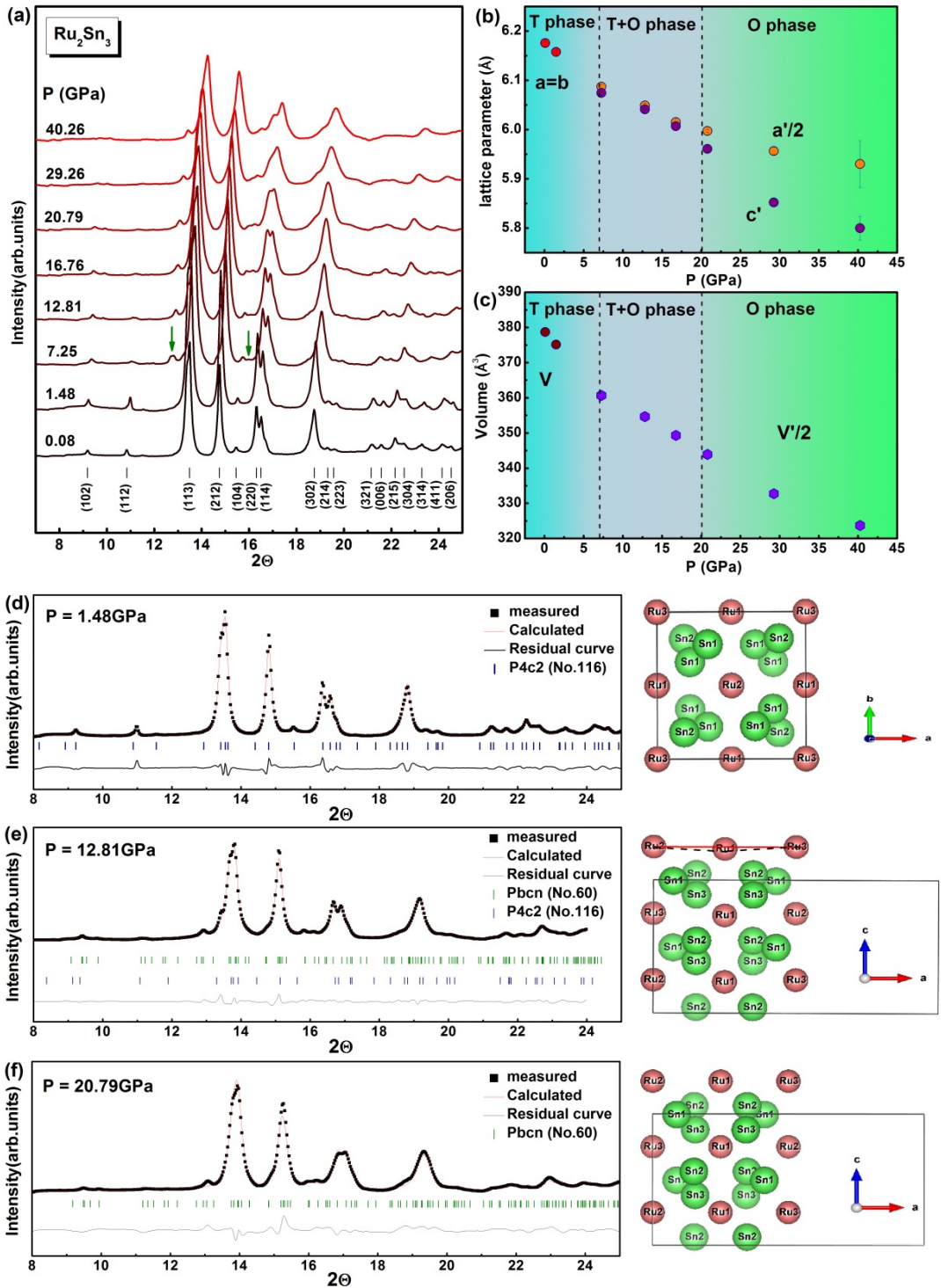

Figure 3 (a) The X-ray diffraction patterns of Ru$_2$Sn$_3$ at various pressures at room temperature. New peaks appear at 7.25 GPa, as indicated by green arrows. (b) The pressure dependence of the lattice parameters. (c) The cell volume as a function of pressure. (d)-(f) Unit cell refinement results for the T phase, the T-O phase mixture, and the O phase, displaying the evolution of crystal structure with pressure.

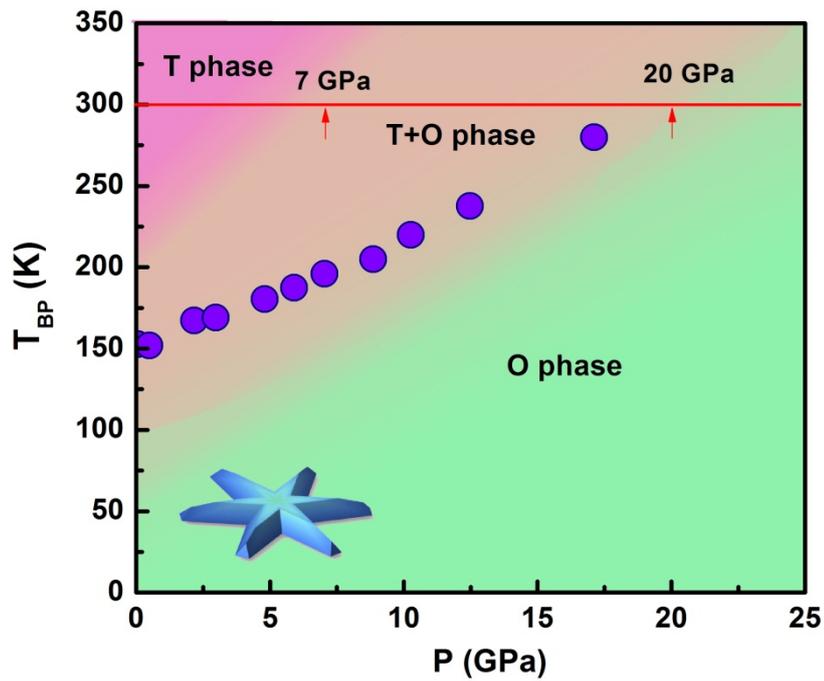

Figure 4 The temperature-pressure phase diagram of Ru$_2$Sn$_3$, showing that the low temperature O phase that displays the non-trivial topological insulating state (represented by the star-shaped pattern of the surface state E vs. k dispersion) can be stabilized to room temperature. The boundaries of the mixed phase are determined through the evolution of T$_1$ and T$_2$ with pressure, as indicated in Fig.2a.